\documentclass[onecolumn,preprintnumbers,eqsecnum,superscriptaddress,12pt]{revtex4}
\usepackage{amsfonts}
\usepackage{amssymb}
\usepackage{amsmath}
\usepackage{epsf}
\usepackage{graphicx}
\usepackage{dcolumn}
\usepackage{bm}

\usepackage{color}

\def\nn{{\nonumber}}

\newcommand{\beq}{\begin{equation}}
\newcommand{\eeq}{\end{equation}}
\newcommand{\beqs}{\begin{eqnarray}}
\newcommand{\eeqs}{\end{eqnarray}}

\newcommand{\be}{\begin{equation}}
\newcommand{\ee}{\end{equation}}
\newcommand{\bea}{\begin{eqnarray}}
\newcommand{\eea}{\end{eqnarray}}

\begin{document}

\title{Hydrodynamics with conserved current via AdS/CFT correspondence in the Maxwell-Gauss-Bonnet gravity }
\author{Ya-Peng Hu}\email{yapenghu@itp.ac.cn}
\address{Center for High-Energy Physics, Peking University, Beijing 100871, China}

\author{Peng Sun}\email{sun_peng_1982@yahoo.com.cn}
\address{Center for High-Energy Physics, Peking University, Beijing 100871, China}

\author{Jian-Hui Zhang}\email{zhangjh@mppmu.mpg.de}
\address{Center for High Energy Physics, Peking University, Beijing 100871, China}

\begin{abstract}
Using the AdS/CFT correspondence, we study the hydrodynamics with conserved current from the dual Maxwell-Gauss-Bonnet gravity. After constructing the perturbative solution to the first order based on the boosted black brane solution in the bulk Maxwell-Gauss-Bonnet gravity, we extract the stress tensor and conserved current of the dual conformal fluid on its boundary, and also find the effect of Gauss-Bonnet term on the dual conformal fluid. Our results show that the Gauss-Bonnet term can affect the parameters such as the shear viscosity $\eta$, entropy density $s$, thermal conductivity $\kappa$ and electrical conductivity $\sigma$. However, it does not affect the so-called Wiedemann-Franz law which relates $\kappa$ to $\sigma$, while it affects the ratio $\eta/s$. In addition, another interesting result is that the $\eta/s$ can also be affected by the bulk Maxwell field in our case, which is consistent with some previous results predicted through the Kubo formula. Moreover, the anomalous magnetic and vortical effects by adding the Chern-Simons term are also considered in our case in the Maxwell-Gauss-Bonnet gravity.


\end{abstract}

\maketitle

\vspace*{1.cm}

\newpage

\section{Introduction}
The AdS/CFT correspondence~\cite{Maldacena:1997re,Gubser:1998bc,Witten:1998qj,Aharony:1999ti} provides a remarkable connection between a gravitational theory and a quantum field theory. According to the correspondence, the gravitational theory in an asymptotically AdS spacetime can be formulated in terms of a quantum field theory on its boundary. Particularly, the dynamics of a classical gravitational theory in the bulk is mapped into the strongly coupled quantum field theory on the boundary. Therefore, AdS/CFT provides a useful tool and some insights to investigate the strongly coupled field theory from the dual classical gravitational theory~\cite{Herzog:2009xv,CasalderreySolana:2011us}.

One of the situations where we need the knowledge of strongly coupled quantum field theory is the quark-gluon plasma (QGP) formed in ultra-relativistic heavy-ion collisions at the Relativistic Heavy Ion Collider (RHIC) and the Large Hadron Collider (LHC). After the collision of the ions, QGP thermalizes rapidly and comes into local thermal equilibrium, and then hadronizes when the local temperature goes down to the deconfinement temperature. Note that, it has been observed that the QGP behaves almost like a perfect fluid in the local thermal equilibrium regime. However, in this regime the perturbation quantum chromodynamics (QCD) breaks down as the the strong QCD coupling remains strong near the deconfinement temperature, while the Lattice QCD is not well-suited in dealing with real time physics and Lorentzian correlation functions. Therefore, an important way to gain some insights into the physics of QGP is through the AdS/CFT correspondence~\cite{CasalderreySolana:2011us}.

Indeed, some early work showed that the AdS/CFT correspondence can be used to describe the hydrodynamical behavior of quantum field via the dual gravity in the bulk. This can be understood from the fact that the hydrodynamics can be viewed as an effective description of an interacting quantum field theory in the long wave-length limit, i.e. when the length scales under consideration are much larger than the correlation length of the quantum field theory. For fluids dual to Einstein gravity in the bulk, the shear viscosity has been
calculated using the Kubo formula in various theories~\cite{Policastro:2001yc,Kovtun:2003wp,Buchel:2003tz,Kovtun:2004de}.
And the ratio of shear viscosity over entropy density $\eta/s$ was
found to be $1/4\pi$ for a large class of conformal field theories
with gravity dual in the large N
limit~\cite{{Mas:2006dy},{Son:2006em},{Saremi:2006ep},{Maeda:2006by},{Cai:2008in}}.
Then it was found that the ratio $\eta/s$ only depends on the value
of effective coupling of transverse gravitons evaluated on the
horizon~\cite{Brustein:2008cg,Iqbal:2008by,Cai:2008ph,Astefanesei:2010dk}. Note that,
this ratio gets only positive corrections from some Large N
effect~\cite{Buchel:2004di,Myers:2008yi}. Therefore, this value was
once conjectured as an universal lower bound for all
materials~\cite{
Policastro:2002se,Buchel:2004qq,Cohen:2007qr,Son:2007vk,Cherman:2007fj,Chen:2007jq,Fouxon:2008pz,Dobado:2008ri,Landsteiner:2007bd}.
However, later studies showed that in higher derivative gravity theories
such as Gauss-Bonnet(GB) gravity the bound is lowered~\cite{Brigante:2007nu,Brigante:2008gz,KP,Brustein:2008cg,Buchel:2009sk,deBoer:2009pn,Camanho:2009vw,Cremonini:2009sy,Ge:2008ni}.
In the case that the dual gravity is the general LoveLock gravity, this lower bound can be further
modification to a more lower bound~\cite{deBoer:2009gx,Camanho:2009hu}.

Recently, the study of the hydrodynamics via dual gravity has been further developed as the Fluid/Gravity
correspondence~\cite{Bhattacharyya:2008jc}. This correspondence provides a more systematic way that maps the boundary fluid to the
bulk gravity. It can be used to construct the stress-energy tensor of the fluid order by order in a derivative expansion from the bulk gravity solution, while the shear viscosity $\eta$, entropy density $s$, and the ratio of the shear viscosity over entropy density
$\eta/s$ can all be calculated from the first order stress-energy tensor~\cite{Hur:2008tq,Erdmenger:2008rm,Banerjee:2008th,Tan:2009yg,Torabian:2009qk,Hu:2010sn}. Besides the stress-energy tensor, this correspondence can also be used to extract the information of the thermal conductivity and electrical conductivity of the boundary fluid from the conserved charge current by adding the Maxwell field in the bulk gravity~\cite{Hur:2008tq,Erdmenger:2008rm,Banerjee:2008th,Tan:2009yg}, where the essential point is that the global U(1) charges in the boundary corresponds to the local U(1) charges in the bulk. In this paper, one of our motivations is to consider the effects of the Gauss-Bonnet term in the bulk on the boundary conserved charge current in contrast to the case of Einstein gravity. After perturbating the boosted black brane solution of the Maxwell-Gauss-Bonnet (MGB) gravity, we calculate the stress-energy tensor and conserved current of the dual conformal fluid to first order based on the first order perturbative gravity solution. Our results show that the Gauss-Bonnet term can affect the parameters of the conformal fluid, such as the shear viscosity $\eta$ , entropy density $s$ , thermal conductivity $\kappa$ and electrical conductivity $\sigma$. However, it does not affect the so-called Wiedemann-Franz law which relates $\kappa$ to $\sigma$, while it affects the ratio $\eta/s$. Another interesting result is that the $\eta/s$ can also been affected by the bulk Maxwell field in MGB gravity, which is consistent with previous work computed through the Kubo formula. In addition, we also consider the anomalous magnetic and vortical effects by adding the Chern-Simons term in our case in the Maxwell-Gauss-Bonnet gravity.

The rest of the paper is organized as follows. In Sec.~II, we briefly review some properties of the Maxwell-Gauss-Bonnet gravity and the black brane solution. In Sec.~III, we construct the perturbative solution to first order. In Sec.~IV, we extract the stress-energy tensor and the conserved current from this first order perturbation solution. And in Sec.~V, we consider the anomalous magnetic and vortical effects from the Chern-Simons term in our case. Sec.~VI is devoted to the conclusion and discussion.


\section{Action and black brane solution of the Maxwell-Gauss-Bonnet gravity}
The action of the $d$ dimensional Maxwell-Gauss-Bonnet gravity with a negative cosmological constant $\Lambda=-(d-2)(d-1)/2\ell^2$ is
\begin{eqnarray}
\label{action} I=\frac{1}{16 \pi G}\int_\mathcal{M}~d^dx \sqrt{-g^{(d)}}
\left(R-2 \Lambda+\alpha L_{GB} \right)-\frac{1}{4g^2}\int_\mathcal{M}~d^dx \sqrt{-g^{(d)}}F^2,
\end{eqnarray}
where $R$ is the Ricci scalar, $\alpha$ with dimension $(length)^2$ is
the GB coefficient and the GB term $L_{GB}$ is
\begin{eqnarray}
\label{LGB} L_{GB} =R^2-4R_{\mu \nu}R^{\mu \nu}+R_{\mu \nu \sigma
\tau}R^{\mu \nu \sigma \tau}.
\end{eqnarray}

For later convenience, set $l=1$ and $16\pi G=1$, the equations of the Maxwell-Gauss-Bonnet gravity are given by
\begin{eqnarray}
\label{eqs}
R_{\mu \nu } -\frac{1}{2}Rg_{\mu \nu}+\Lambda g_{\mu \nu}+\alpha H_{\mu \nu}-\frac{1}{2g^2}\left(F_{\mu \sigma}{F_{\nu }}^{\sigma}-\frac{1}{4}g_{\mu \nu}F^2\right)&=&0~, \\
\nabla_{\mu} {F^{\mu}}_{\nu}&=&0.~~
\end{eqnarray}
where
\begin{equation}
\label{Hmn}
H_{\mu \nu}=2(R_{\mu \sigma \kappa \tau }R_{\nu }^{\phantom{\nu}%
\sigma \kappa \tau }-2R_{\mu \rho \nu \sigma }R^{\rho \sigma
}-2R_{\mu
\sigma }R_{\phantom{\sigma}\nu }^{\sigma }+RR_{\mu \nu })-\frac{1}{2}%
L_{GB}g_{\mu \nu }  ~.
\end{equation}

There have been many exact solutions found~\cite{Des,Whee,Myers,Cai1,Cai:2001dz,Cremonini:2008tw,Nojiri:2001aj,Cvetic:2001bk,Anninos:2008sj}, and we are interested in the $5$-dimensional charged black brane solution~\cite{Cvetic:2001bk,Anninos:2008sj}
\begin{eqnarray}
ds^2=\frac{dr^2}{r^2f(r)}+\frac{r^2}{\ell_c^2}
 \left(\mathop\sum_{i=1}^{3}dx_i^2 \right)-r^2f(r) dt^2, \label{Solution}
\end{eqnarray}%
where
\begin{eqnarray}
\label{f-BH}
f(r) &=& \frac{1}{4\alpha}
 \bigg (
 1-\sqrt{1-8\alpha (1-\frac{2M}{r^{4}}+\frac{Q^2}{r^6})}~
 \bigg), \\
F &=& g\frac{2\sqrt 3 Q}{r^3}dt \wedge dr. ~~
\end{eqnarray}%
Note that, $\ell_c$ is the effective radius of the AdS spacetime in GB gravity, and its expression in $5$-dimensional case is
\begin{eqnarray}
\label{lc}
\ell_c=\sqrt{\frac{1+U}{2}},~~{\rm~~with~~~~}U=\sqrt{1-8\alpha}.
\end{eqnarray}

From~(\ref{Solution}), we easily find that the horizon of the black brane is located at $r=r_{+}$, where $r_{+}$ is the largest root of $f(r)=0$. The Hawking temperature and entropy density are
\begin{eqnarray}
T&=&\frac{(r^2f(r))'}{4 \pi}|_{r=r_{+}}=\frac{1}{2 \pi r_{+}^3}(4M-\frac{3Q^2}{r_{+}^2}),\label{Temperature}\\
s&=&\frac{r_{+}^3}{4G \ell_c^3}.\label{entr}
\end{eqnarray}
In addition, this solution rewritten in the Eddington-Finkelstin coordinate system is
\begin{eqnarray}\label{Solution1}
ds^2 &=& - r^2 f(r)dv^2 + 2 dv dr + \frac{r^2}{\ell_c^2}(dx^2 +dy^2 +dz^2),\\
F &=& g\frac{2\sqrt 3 Q}{r^3}dv \wedge dr. \notag~~
\end{eqnarray}
where $v=t+r_*$ with $dr_*=dr/(r^2f)$.
Moreover, in the boosted frame, it can be written as
\begin{eqnarray}\label{rnboost}
ds^2 &=& - r^2 f(r)( u_\mu dx^\mu )^2 - 2 u_\mu dx^\mu dr + \frac{r^2}{\ell_c^2}P_{\mu \nu} dx^\mu dx^\nu, \\
F &=& -g\frac{2\sqrt 3 Q}{r^3} u_\mu dx^\mu \wedge dr,~~~A=(e A_{\mu}^{ext}-\frac{\sqrt 3 g Q}{r^2}u_{\mu})dx^{\mu}. \notag~~
\end{eqnarray}
with
\begin{equation}
u^v = \frac{1}{ \sqrt{1 - \beta_i^2} }~~,~~u^i = \frac{\beta_i}{ \sqrt{1 - \beta_i^2} },~~P_{\mu \nu}= \eta_{\mu\nu} + u_\mu u_\nu~~.
\end{equation}
where velocities $\beta^i $, $M$, $Q$ and $A_{\mu}^{ext}$ are constants, $x^\mu=(v,x_{i})$ are the boundary coordinates, $P_{\mu \nu}$ is the projector onto spatial directions, and the indices in the boundary are raised and lowered with the Minkowsik metric $\eta_{\mu\nu}$.


\section{The first order perturbative solution}
In this section, we construct the perturbative solution based on the boosted black brane~(\ref{rnboost}). The reason is that we can extract the viscous information of the conformal fluid from this perturbative solution, as can be seen in the following in detail. To procedure this perturbation, we lift the above constant parameters $\beta^i $, $M$, $Q$, $A_{\mu}^{ext}$ to be slowly-varying functions of the boundary coordinates $x^\mu=(v,x_{i})$. Therefore, the metric~(\ref{rnboost}) will be not a solution of the equations of motion~(\ref{eqs}) any more, and we need to add correction terms to make the new metric as a solution. Before discussing these correction terms, we first define the following tensors
\begin{eqnarray}
\label{Tensors}
&&W_{IJ} = R_{IJ} + 4g_{IJ}+\frac{1}{6}\alpha L_{GB} g_{IJ}+\alpha H_{IJ}+\frac{1}{2g^2}\left(F_{IK}{F^{K}}_J +\frac{1}{6}g_{IJ}F^2\right),\\
&&W_I = \nabla_J {F^J}_I~~.
\end{eqnarray}
When we take the parameters as functions of $x^\mu$ in~(\ref{rnboost}), $W_{\mu \nu}$ and $W_{\mu}$ will be nonzero and proportional to the derivatives of the parameters. Therefore, these terms can be considered as the source terms $S_{\mu \nu }$ and $S_{\mu}$, which are canceled by the correction terms. More details, let the parameters expanded around $x^\mu=0$ to first order
\begin{eqnarray}
\beta_i&=&\partial_{\mu} \beta_{i}|_{x^\mu=0} x^{\mu},~~~M=M(0)+\partial_{\mu} M|_{x^\mu=0} x^{\mu},~~~Q=Q(0)+\partial_{\mu} Q|_{x^\mu=0} x^{\mu},\notag\\
A_{\mu}^{ext}&=&A_{\mu}^{ext}(0)+\partial_{\nu} A_{\mu}^{ext}|_{x^\mu=0} x^{\nu}. \label{Expand}
\end{eqnarray}
where we have assumed $\beta^i(0)=0$. After inserting the metric (\ref{rnboost}) with~(\ref{Expand}) into $W_{\mu \nu }$ and $W_{\mu}$, the first order source terms can be $S^{(1)}_{\mu \nu } = - W_{\mu \nu }$ and $S^{(1)}_{\mu} = - W_{\mu}$. Therefore, after fixing some gauge and considering the spatial $SO(3)$ symmetry preserved in the background metric~(\ref{Solution1}), the choice for the first order correction terms around $x^\mu=0$ can be
\begin{eqnarray}\label{correction}
&&{ds^{(1)}}^2 = \frac{ k(r)}{r^2}dv^2 + 2 h(r)dv dr + 2 \frac{j_i(r)}{r^2}dv dx^i +\frac{r^2}{\ell_{c}^2} \left(\alpha_{ij} -\frac{2}{3} h(r)\delta_{ij}\right)dx^i dx^j, \\
&&A^{(1)} = a_v (r) dv + a_i (r)dx^i~~.
\end{eqnarray}
Note that, for gauge field part, $a_r(r)$ does not contribute to field strength, thus the choice $a_r(r)=0$ is trivial.
Therefore, the first order perturbation solution can be obtained from the vanishing $W_{\mu \nu } = (\text{effect from correction}) - S^{(1)}_{\mu \nu }$ and $W_{\mu} = (\text{effect from correction}) - S^{(1)}_{\mu}$.

For the Maxwell equation, what we have to do is just solving following equations
\begin{eqnarray}\label{Maxwell}
&&W_v = \frac{f(r)}{r}\left\{ r^3 {a_v}' (r) + 4\sqrt 3 g Q h(r) \right\}'- S^{(1)}_{v}(r)=0~,\\\nn
&& W_r = - \frac{1}{r^3}\left\{  r^3 {a_v}'(r) +4\sqrt 3 g Q h(r)  \right\}'-S^{(1)}_r (r)=0~,\\\nn
&&W_i = \frac{1}{r} \left\{ r^3 f(r) {a_i}'(r) - \frac{2 \sqrt 3 g Q }{r^4} j_i (r) \right\}' -  S^{(1)}_i (r)=0~.
\end{eqnarray}
where
\begin{eqnarray}
&&S_v^{(1)}(r)=g\frac{2 \sqrt{3}}{r^3} \left(\partial _vQ+Q \partial _i\beta _i\right),\\\nn&&S_r^{(1)}(r)=0,\\\nn&&S_i^{(1)}(r)=g \left(-\frac{\sqrt{3}}{r^3} \left(\partial _iQ+Q \partial _v\beta _i\right)-\frac{1}{r} \frac{e}{g}F^{\rm ext}_{vi} \right)~.
\end{eqnarray}
and $F^{\rm ext}_{vi}\equiv \partial_v A^{\rm ext}_i-\partial_i A^{\rm ext}_v$ is the external field strength tensor,
prime means derivative of $r$ coordinate.

For equations of gravity, they are complicated, which can be seen more details in the appendix~\ref{A}. By solving all the above equations, several coefficients of the first order correction terms are
\begin{eqnarray}
&&h(r) = 0,~~ k(r) = \frac{2}{3}r^3 \partial_i\beta^i,~~a_{v}(r)=0,\\\nn &&\alpha_{ij} =  \alpha(r)\left\{ (\partial_i \beta_j + \partial_j \beta_i )-\frac{2}{3} \delta_{ij}\partial_k \beta^k \right\},
\end{eqnarray}
where $\alpha(r)$ and its asymptotic expression are
\begin{eqnarray}
\alpha(r)&&= \int_{\infty }^{r}\frac{s^{3}-2\alpha s^{2}[s^{2}f(s)]^{^{\prime}}-(r_{+}^3-2\alpha r_{+}^2(r^2f)'|_{r_{+}})}{-s+2\alpha [ s^{2}f(s)]^{^{\prime }}}\frac{1}{%
s^{4}f(s)}ds \notag\\
&&\approx \frac{\ell_c^2}{r}-\frac{1}{r^4}\frac{\alpha(r_{+}^6+12Q^2 \alpha -16Mr_{+}^2 \alpha)}{r_{+}^3(1-\sqrt{1-8\alpha})\sqrt{1-8\alpha}}+O(\frac{1}{r})^5.
\end{eqnarray}
Since the remaining equations $W_i =0$ and $W_{ri} =0$ are coupled to each other, it is more difficult to solve them. These equations are
\begin{eqnarray}
&&\frac{r}{2} \left(\frac{j_i'(r)}{r^3}\right)'-\frac{\sqrt{3} Q}{g r^3} a_i'(r)+\frac{8\alpha j_i(r)f'(r)}{r^3}+\frac{6\alpha j_i'(r)f(r)}{r^3}- \frac{2\alpha j_i'(r)f'(r)}{r^2}-\frac{2\alpha j_i''(r)f(r)}{r^2}=S_{r i}^{(1)}(r), \notag\\
&& \left(r^3 f(r) a_i'(r)-\frac{2 \sqrt{3} g Q}{r^4} j_i(r)\right)'=r S_i^{(1)}(r). \label{appeq1}
\end{eqnarray}
For the exact solutions to these equations, see appendix \ref{B}. Note that, we can decompose $j_{i}(r)$ and $a_{i}(r)$ as
\begin{eqnarray}
a_i(r) &=& a_\beta(r) \partial_v \beta_i + a_Q(r) (\partial_i Q+Q\partial_v \beta_i) +a_F(r) F^{\rm ext}_{vi}, \\
j_i(r) &=& j_\beta(r) \partial_v \beta_i + j_Q(r) (\partial_i Q+Q\partial_v \beta_i) +j_F(r) F^{\rm ext}_{vi}.
\end{eqnarray}
With this decomposition, the gauge field can be written as a covariant form
\begin{eqnarray}
A^{(1)}_\mu(r)=a_\beta(r) u^\nu \partial_\nu u_\mu +a_Q(r) u^\nu F^{(Q)}_{\nu\mu} +a_F(r)u^\nu F^{\rm ext}_{\nu\mu},
\end{eqnarray}
where $F_{\lambda  \nu }^{(Q)}\equiv \partial _{\lambda }\left(Q u_{\nu }\right)-\partial _{\nu }\left(Q u_{\lambda }\right)$. In addition, there are some relations between these equations
\begin{eqnarray}
 &&W_v + r^2 f(r)W_r =0 ~:~ S_v^{(1)} + r^2 f(r)S_r^{(1)} =0, \notag \\
 &&W_{vi} + r^2 f(r) W_{ri} =0 ~:~ S_{vi} + r^2 f(r) S_{ri} = 0, \notag \\
 &&W_{vv} + r^2 f(r)W_{vr} =0 ~:~ S_{vv} + r^2 f(r) S_{vr} = 0. \label{constraint}
 \end{eqnarray}
which can be considered as the constraint equations.
In our paper, after using the first order source terms in the appendix~\ref{A}, we can further rewrite the constrain equations~(\ref{constraint}) as
\begin{eqnarray}
&&3 \partial _vM+4 M \partial _i\beta _i=0,  \\\nn &&\partial _iM+4 M \partial _v\beta _i=\sqrt{3} Q   \frac{e}{g}F^{\rm ext}_{vi},\\\nn &&\partial _vQ+Q \partial _i\beta _i=0~~.
\end{eqnarray}
In the later, we can see that these equations can be expressed as a covariant form, which are nothing but the exact conservation equations of the zeroth order stress-energy tensor and conserved current.

Therefore, after adding the correction terms, the first-order metric expanded in boundary derivatives around $x^{\mu}=0$ is given explicitly as
\begin{eqnarray}
ds^{2}&=&2dvdr-r^{2}f(M_{0},Q_{0},r)dv^{2}+\frac{r^{2}}{\ell_c^2}dx_{i}^{2}-r^{2}x^{\mu }C_{1}(r)\partial _{\mu }Mdv^{2}-2x^{\mu }\partial _{\mu }\beta_{i}dx^{i}dr \notag\\
& &-2x^{\mu}r^{2}[\frac{1}{\ell_c^2}-f(M_{0},Q_{0},r)]\partial _{\mu }\beta _{i}dx^{i}dv-r^{2}x^{\mu }C_{2}(r)\partial _{\mu }Qdv^{2}
+2r^2\alpha(r)\sigma_{ij}dx^idx^j\notag\\
&&+\frac{2}{3}r\partial_i\beta^idv^2+2\frac{j_{i}(r)}{r^2}dvdx^i,\label{GloballySolution1}
\end{eqnarray}
where
\begin{eqnarray}
f(M_{0},Q_{0},r)&=&f(M(x^{\mu}),Q(x^{\mu}),r)|_{x^{\mu}=0},~~C_{1}(r)=\frac{\partial f(M(x^{\mu}),Q(x^{\mu}),r)}{\partial M}|_{x^{\mu}=0},\notag\\
C_{2}(r)&=&\frac{\partial f(M(x^{\mu}),Q(x^{\mu}),r)}{\partial Q}|_{x^{\mu}=0},~~\sigma_{ij}=\partial_{(i} \beta_{j)}-\frac{1}{3} \delta_{ij}\partial_k \beta^k.
\end{eqnarray}
From which the global first-order metric can be constructed in a covariant form
\begin{eqnarray}
ds^2&=&-r^2 f(r) u_{\mu } u_{\nu } dx^{\mu } dx^{\nu }-2 u_{\mu } dx^{\mu } dr+\frac{r^2}{\ell_c^2} P_{\mu  \nu } dx^{\mu } dx^{\nu }+\Bigg[\frac{2 r}{3} u_{\mu } u_{\nu } \partial _{\lambda }u^{\lambda }\\\nn& & - \frac{2}{r^2} \left\{
\frac{1}{2}j_{\beta } (r) u^{\lambda } \partial _{\lambda }\left(u_{\mu } u_{\nu }\right)+  j_Q (r)  u_{\mu } u^{\lambda } F _{\lambda  \nu }^{(Q)}+  j_{F}(r) u_{\mu } u^{\lambda } F^{\rm ext}_{\lambda  \nu } \right\}\\\nn && +2 r^2 \alpha (r) \sigma _{\mu  \nu }\Bigg] dx^{\mu } dx^{\nu }~~,\label{GloballySolution}
\end{eqnarray}
where we have took the covariant expression and the definitions of $\sigma_{\mu \nu}$
\begin{eqnarray}
\sigma ^{\mu  \nu }\equiv \frac{1}{2} P^{\mu  \alpha } P^{\nu  \beta } \left(\partial _{\alpha }u_{\beta }+\partial _{\beta }u_{\alpha }\right)-\frac{1}{3} P^{\mu  \nu } \partial _{\alpha }u^{\alpha }~~.
\end{eqnarray}

The chemical potential is defined as
\begin{eqnarray}\label{chemical potential}
\mu = A_v (r_+) - A_v (\infty) ~~.
\end{eqnarray}
Using the same discussion in reference \cite{Hur:2008tq},  we can find that its first order expression is
 \begin{eqnarray}
 \mu = 
 \frac{\sqrt 3 gQ(x)}{  r_+ ^2 (x)}~~.
 \end{eqnarray}
which keeps the same expression but here $Q$ and $r_+$ are not constants.

\section{The Stress tensor and conserved current of conformal field via dual gravity}
Basing on the above perturbative solution, we now extract the information of the dual conformal field. And we first discuss its stress tensor $\tau _{\mu\nu}$ which can be obtained through the following relationship~\cite{Myers:1999psa}
\begin{eqnarray}\label{relation}
\label{Tik-CFT} \sqrt{-h}h^{\mu\nu}<\tau _{\nu\sigma}>=\lim_{r\rightarrow
\infty }\sqrt{-\gamma }\gamma ^{\mu\nu}T_{\nu\sigma}.
\end{eqnarray}
where $h_{\mu\nu}$ is the background metric upon which the dual field theory resides, $\gamma ^{\mu\nu}$ is the boundary metric obtained from
the well-known ADM decomposition
\begin{eqnarray}
ds^2 = \gamma_{\mu\nu}(dx^\mu + V^\mu dr)(dx^\nu + V^\nu dr) + N^2 dr^2~~,
\end{eqnarray}
and $T_{ab}$ is the corresponding boundary stress tensor which is defined
\begin{equation}
T_{ab}=\frac{2}{\sqrt{-\gamma }}\frac{\delta }{\delta \gamma
^{ab}}\left( I+I_{\mathrm{sur}}+I_{\text{ct}}^0 \right),
\label{Tab}
\end{equation}
where
\begin{equation}
I_{\mathrm{sur}}=-\frac{1}{8\pi G}\int_{\partial
\mathcal{M}}d^{4}x\sqrt{-\gamma }K-\frac{\alpha}{4\pi G}\int_{\partial \mathcal{M}}d^{4}x\sqrt{-\gamma }%
 \left( J-2{\rm G}_{ab} K^{ab}\right) ~, \label{Surfaceterm}
\end{equation}
is the generalized surface term of the 5-dimensional GB gravity and we have recovered the gravitational coupling hereafter~\cite{Myers:1987yn,Brihaye:2008xu}, $K$ is
the trace of the extrinsic curvature of the boundary, ${\rm G}_{ab}$ is the Einstein tensor of the metric $\gamma _{ab}$ and $J$ is the
trace of the tensor
\begin{equation}
J_{ab}=\frac{1}{3}%
(2KK_{ac}K_{b}^{c}+K_{cd}K^{cd}K_{ab}-2K_{ac}K^{cd}K_{db}-K^{2}K_{ab})~.
\label{Jab}
\end{equation}
In addition,
\begin{eqnarray}
\label{Lagrangianct} I_{\mathrm{ct}}^0 &=&\frac{1}{8\pi
G}\int_{\partial \mathcal{M}} d^{4}x\sqrt{-\gamma } \left[
  -\frac{2+U}{\ell_c }
 -\frac{\ell_c }{4}(2-U)\mathsf{R}\right].
\end{eqnarray}
is the corresponding boundary counterterm and  $\mathsf{R}$ is the
curvature scalar associated with the
induced metric on the boundary $\gamma_{ab}$~\cite{Brihaye:2008xu,Cremonini:2009ih}. Note that, it is obviously seen that the above boundary counterterm can recover the known counterterm expression in the Einstein gravity when $\alpha \to 0$~\cite{Balasubramanian:1999re,Emparan:1999pm,Mann:1999pc}.

From~(\ref{Tab}), the boundary stress-energy tensor is
\begin{equation}
 T_{ab}=\frac{1}{8 \pi G}[K_{ab}-\gamma _{ab}K
 +2\alpha (Q_{ab}-\frac{1}{3}Q\gamma_{ab})
-\frac{2+U}{\ell_c}\gamma _{ab} +\frac{\ell_c}{2}(2-U)%
( \mathsf{R}_{ab}-\frac{1}{2}\gamma _{ab}\mathsf{R})], \label{TabCFT}
\end{equation}
where $Q_{ab}$ is
\begin{eqnarray}
&Q_{ab}= 2KK_{ac}K^c_b-2 K_{ac}K^{cd}K_{db}+K_{ab}(K_{cd}K^{cd}-K^2)
+2K \mathsf{R}_{ab}+\mathsf{R}K_{ab} -2K^{cd}\mathsf{ R}_{cadb}-4
\mathsf{R}_{ac}K^c_b.~{~~~~~~} \notag
\end{eqnarray}

Therefore, after inserting the explicit metric~(\ref{GloballySolution1}) into~(\ref{TabCFT}), the corresponding non-zero components of the boundary energy-momentum tensor are
\begin{eqnarray}
T_{vv}&=&\frac{3M}{8\pi G r^2\ell_{c}},~T_{ij}=-\frac{r_{+}^6+(12Q^2-16Mr_{+}^2 \alpha)}{8\pi G r^2 \ell_{c}} \sigma_{ij}=-\frac{r_{+}^2(r_{+}-8 \pi T \alpha)}{8\pi G r^2 \ell_{c}} \sigma_{ij}~(i\neq j),\notag\\
~T_{ij}&=&\frac{1}{16\pi G}[\frac{2M}{r^2 \ell_{c}}-\frac{2r_{+}^2(r_{+}-8 \pi T \alpha)}{ r^2 \ell_{c}} \sigma_{ij}]~(i=j). \label{BoundaryST}
\end{eqnarray}

The background metric upon which the dual field theory resides can be obtained from~\cite{Myers:1999psa}
$h_{ab}=\lim_{r \rightarrow \infty}
\frac{\ell^2_c}{r^2}\gamma_{ab}$, and
\begin{equation}
ds^2=h_{ab}dx^adx^b=-dv^2+dx^2+dy^2+dz^2.
\end{equation}
Therefore, according to~(\ref{relation}) and~(\ref{BoundaryST}), the expectation value of the
first order stress tensor of the dual theory $\tau _{\mu \nu}$ corresponding to the global metric~(\ref{GloballySolution}) is
\begin{equation}
\tau_{\mu \nu}=\frac{1}{16\pi G}[\frac{2M}{\ell_{c}^3}(\eta_{\mu\nu} + 4 u_\mu u_\nu)-\frac{2r_{+}^2(r_{+}-8 \pi T \alpha)}{ \ell_{c}^3}\sigma_{\mu\nu}]=P(\eta_{\mu\nu} + 4 u_\mu u_\nu  ) - 2 \eta \sigma_{\mu\nu}. \label{StressTensor}
\end{equation}
where we have rewritten it as a covariant form. It should be emphasized that we can also obtain~(\ref{StressTensor}) by directly inserting the global metric~(\ref{GloballySolution}) into~(\ref{TabCFT}). From~(\ref{StressTensor}), the pressure and viscosity are read off
\begin{equation}
P=\frac{M}{8 \pi G \ell_{c}^3},~~~\eta=\frac{r_{+}^2(r_{+}-8 \pi T \alpha)}{16 \pi G \ell_{c}^3}. \label{ets}
\end{equation}
and the entropy density $s$ can be computed through
\begin{equation}
s=\frac{\partial P}{\partial T}=\frac{r_{+}^3}{4G \ell_{c}^3}. \label{entr1}
\end{equation}
The ratio of $\eta$ and $s$ can also be found from (\ref{ets}) (\ref{entr}) that
\begin{equation}
\label{ror}
 {\eta \over s} = {1 \over 4 \pi} (1-\frac{8 \pi T \alpha}{r_{+}}).
\end{equation}
which is consistent with the previous result using the Kubo formula~\cite{Cai:2008ph,Myers:2009ij}.
In addition, from $\tau _{\mu \nu}$, we can obtain the zeroth order energy-momentum tensor
\begin{equation}
\tau_{(0)}^{\mu\nu} =\frac{M}{8 \pi G \ell_{c}^3} (\eta^{\mu\nu} + 4 u^\mu u^\nu),
\end{equation}

On the other hand, the boundary current can be computed via
\begin{equation}\label{current}
J^\mu = \lim_{r \rightarrow \infty} \frac{r^4}{\ell_{c}^4} \frac{1}{\sqrt{-\gamma}} \frac{\delta S_{cl}}{\delta \tilde A_\mu} =  \lim_{r \rightarrow \infty}  \frac{r^4}{\ell_{c}^4} \frac{N}{g^2} F^{r \mu}~~,
\end{equation}
where $\tilde A_\mu$ is the gauge field which is projected to the boundary.
The current is
\begin{eqnarray}
J^\mu &=& J_{(0)}^\mu + J_{(1)}^\mu, \\ \label{first order current}
J_{(1)}^{\mu }&=& \frac{1}{g \ell_{c}} \left\{-2 \sqrt{3} Q \frac{j_{\beta }\left(r_+\right)}{r_+^4} u^{\lambda } \partial _{\lambda }u^{\mu }+\left(-2 \sqrt{3} Q \frac{j_Q\left(r_+\right)}{r_+^4}-\frac{\sqrt{3}}{r_+}\right) u^{\lambda } F ^{(Q)}{}_{\lambda }{}^{\mu } \right.\nonumber\\
&& \left.+\left(-2 \sqrt{3} Q \frac{j_F\left(r_+\right)}{r_+^4}+\frac{e}{g}r_+\right) u^{\lambda } F ^{\rm ext}{}_{\lambda }{}^{\mu }\right\},
\end{eqnarray}
where the zeroth order boundary (particle number) current is
\begin{equation}
J_{(0)}^{\mu} = \frac{2\sqrt 3 Q}{g \ell_{c}^3} u^\mu :=nu^\mu.
\end{equation}
and $j_\beta(r_+)$, $j_Q(r_+)$, $j_F(r_+)$ are values of each function at the horizon, which can be read from (\ref{j at rp}) and found same as the reference~\cite{Hur:2008tq}
\begin{eqnarray}
\frac{j_{\beta }\left(r_+\right)}{r_+^4}=\frac{2 \left(2 r_+^6+Q^2\right)}{8 M r_+^3}~~,~~\frac{j_Q\left(r_+\right)}{r_+^4}=-\frac{Q}{8 M r_+^3}~~,
~~\frac{j_{F}\left(r_+\right)}{r_+^4}=-\frac{e}{g}\frac{\sqrt{3} Q}{8 M r_+}~~
\label{jis}.
\end{eqnarray}
Thus after a little work, the constraint equation~(\ref{constraint}) can be expressed covariantly as
\begin{eqnarray}\label{conservation}
&&\partial_\mu T_{(0)}^{\mu\nu} = 2\sqrt 3  Q \frac{e}{g}F_{\rm ext}^{ \mu\nu}u_\mu~, \\ \nn&&
\partial_\mu J_{(0)}^\mu = 0~~.
\end{eqnarray}
which are just the exact zeroth order conservation equations.

After some algebra, the current can be simplified as
 \begin{eqnarray}\label{familiar current}
 J_{(1)}^\mu  &=& \frac{1}{\ell_{c}}\big(- \frac{\pi ^2 T^3 r_+^7}{4g^2 M^2}  P^{\mu  \nu } \partial _{\nu }\frac{\mu }{T}
+ \frac{\pi ^2 T^2 r_+^7}{4g^2 M^2} u^{\lambda } eF ^{\rm ext}{}_{\lambda }{}^{\mu })~~,\label{cur1}\\
 &:=& -\kappa  P^{\mu  \nu } \partial _{\nu }\frac{\mu }{T}+\frac{\sigma}{e} u^{\lambda } F ^{\rm ext}{}_{\lambda }{}^{\mu }~~,
\end{eqnarray}
where we have used the following relations to obtain(\ref{cur1})
 \begin{eqnarray}
&&u^{\lambda } \partial _{\lambda }u^{\mu }=\frac{\sqrt{3} Q}{4
M}\frac{e}{g} u^{\lambda } F_{\lambda  \mu }^{\text{ext}}-\frac{1}{4
M} P^{\mu  \nu } \partial _{\nu }M~,\\
&&Q=\frac{\mu  r_+^2}{\sqrt{3} g},\\
&&r_+^2 r_-^2+r_-^4=\frac{\mu ^2 r_+^2}{3 g^2}=2 M-r_+^4,\\
&&T=\frac{1}{2 \pi  r_+^3} \left(2 r_+^4-r_+^2
r_-^2-r_-^4\right)=\frac{6 r_+^2-\mu ^2/g^2}{6 \pi  r_+}=\frac{3
r_+^4-2 M}{2 \pi  r_+^3}~.
\end{eqnarray}

Therefore, the coefficient of thermal conductivity $\kappa $  and the
electrical conductivity can be read off
\begin{eqnarray} \label{TEconductivity}
\kappa =  \frac{\pi ^2 T^3 r_+^7}{4 g^2M^2 \ell_{c}}, \;\;\quad
\sigma =\frac{\pi ^2 e^2T^2 r_+^7}{4g^2 M^2 \ell_{c}}.
 \end{eqnarray}
Obviously, the so-called Wiedermann-Franz law relating the thermal conductivity and electrical conductivity holds~\cite{Hur:2008tq}
\begin{equation}
\kappa=\sigma T/e^2.
\end{equation}

\section{Anomalous magnetic and vortical effects from the Chern-Simons term} \label{C}
In this section, we will deduce the anomalous magnetic and vortical effects by adding the Chern-Simons term in our case in the Maxwell-Gauss-Bonnet gravity. The Chern-Simons term is added in the action (\ref{action}) such that
\begin{equation}
\label{action1} I=\frac{1}{16 \pi G}\int_\mathcal{M}~d^dx \sqrt{-g^{(d)}}
\left(R-2 \Lambda+\alpha L_{GB} \right)-\frac{1}{4g^2}\int_\mathcal{M}~d^dx \sqrt{-g^{(d)}}(F^2+\frac{4\kappa_{cs} }{3}\epsilon ^{\mu \nu \rho \sigma \tau }A_{\mu}F_{\nu \rho}F_{\sigma \tau }),
\end{equation}
and the new equations of the Maxwell-Gauss-Bonnet gravity are
\begin{eqnarray}
\label{eqs1}
R_{\mu \nu } -\frac{1}{2}Rg_{\mu \nu}+\Lambda g_{\mu \nu}+\alpha H_{\mu \nu}-\frac{1}{2g^2}\left(F_{\mu \sigma}{F_{\nu }}^{\sigma}-\frac{1}{4}g_{\mu \nu}F^2\right)&=&0~, \\
\nabla_{B} {F^{B}}_{A}-\kappa_{cs} \epsilon_{ABCDE}F^{BC}F^{DE} &=&0.\nonumber~~
\end{eqnarray}
Note that, the Chern-Simons term does not effect the expressions of solutions, therefore (\ref{Solution}) is also the the solution of equations (\ref{eqs1}). The difference is that we should define the new following tensors
\begin{eqnarray}
\label{Tensors}
&&W_{IJ} = R_{IJ} + 4g_{IJ}+\frac{1}{6}\alpha L_{GB} g_{IJ}+\alpha H_{IJ}+\frac{1}{2g^2}\left(F_{IK}{F^{K}}_J +\frac{1}{6}g_{IJ}F^2\right),\\
&&W_A = \nabla_{B} {F^{B}}_{A}-\kappa_{cs} \epsilon_{ABCDE}F^{BC}F^{DE}~~.
\end{eqnarray}
Therefore, they are same for equations of gravity, while for the the Maxwell equations they are
\begin{eqnarray}\label{Maxwell}
&&W_v = \frac{f(r)}{r}\left\{ r^3 {a_v}' (r) + 4\sqrt 3 g Q h(r) \right\}'- S^{(1)}_{v}(r)=0~,\\\nn
&& W_r = - \frac{1}{r^3}\left\{  r^3 {a_v}'(r) +4\sqrt 3 g Q h(r)  \right\}'-S^{(1)}_r (r)=0~,\\\nn
&&W_i = \frac{1}{r} \left\{ r^3 f(r) {a_i}'(r) - \frac{2 \sqrt 3 g Q }{r^4} j_i (r) \right\}' -  S^{(1)}_i (r)=0~.
\end{eqnarray}
where
\begin{eqnarray}
&&S_v^{(1)}(r)=g\frac{2 \sqrt{3}}{r^3} \left(\partial _vQ+Q \partial _i\beta _i\right),\\\nn&&S_r^{(1)}(r)=0,\\\nn&&S_x^{(1)}(r)=g \left(-\frac{\sqrt{3}}{r^3} \left(\partial _xQ+Q \partial _v\beta _x\right)-\frac{1}{r} \frac{e}{g}F^{\rm ext}_{vx} \right)-\kappa_{cs}\frac{16g\ell_{c}Q}{r^6}\left(\sqrt{3}er^2F^{\rm ext}_{zy}-3gQ\partial _z\beta _y+3gQ\partial _y\beta _z \right)~.
\end{eqnarray}
and $F^{\rm ext}_{vi}\equiv \partial_v A^{\rm ext}_i-\partial_i A^{\rm ext}_v$ is the external field strength tensor,
prime means derivative of $r$ coordinate.

By considering the Chern-Simons term, the boundary current can be computed via
\begin{equation}\label{current}
J^\mu = \lim_{r \rightarrow \infty} \frac{r^4}{\ell_{c}^4} \frac{1}{\sqrt{-\gamma}} \frac{\delta S_{cl}}{\delta \tilde A_\mu} =  \lim_{r \rightarrow \infty}  \frac{r^4}{\ell_{c}^4} \frac{N}{g^2} (F^{r \mu}+\frac{4\kappa_{cs} }{3}\epsilon ^{r\mu \rho \sigma \tau }A_{\rho
}F_{\sigma \tau })~~,
\end{equation}
where $\tilde A_\mu$ is the gauge field which is projected to the boundary.
After some algebra, the current is
\begin{eqnarray}
J^\mu &=& J_{(0)}^\mu + J_{(1)}^\mu, \\ \label{first order current}
J_{(1)}^{\mu}&=& \frac{1}{g \ell_{c}} \left\{-2 \sqrt{3} Q \frac{j_{\beta }\left(r_+\right)}{r_+^4} u^{\lambda } \partial _{\lambda }u^{\mu }+\left(-2 \sqrt{3} Q \frac{j_Q\left(r_+\right)}{r_+^4}-\frac{\sqrt{3}}{r_+}\right) u^{\lambda } F ^{(Q)}{}_{\lambda }{}^{\mu } \right.\nonumber\\
&& \left.+\left(-2 \sqrt{3} Q \frac{j_F\left(r_+\right)}{r_+^4}+\frac{e}{g}r_+\right) u^{\lambda } F ^{\rm ext}{}_{\lambda }{}^{\mu }\right\}+\kappa_{cs} \left(\frac{\sqrt{3}eQB^{\mu}}{gMr_{+}^4}(Q^2+4r_{+}^6)+\frac{6Q^2}{M}\omega ^{\mu }\right),\nonumber\\
\end{eqnarray}
where we have assumed $A_{\mu}^{ext}(0)=0$ in (\ref{Expand}), and
\begin{equation}
J_{(0)}^{\mu} = \frac{2\sqrt 3 Q}{g \ell_{c}^3} u^\mu :=nu^\mu,~~B^{\mu }=\frac{1}{2}\epsilon ^{\mu \nu \rho \sigma}u_{\nu}F_{\rho \sigma}^{ext},~~\omega ^{\mu }=\frac{1}{2}\epsilon ^{\mu \nu \rho \sigma }u_{\nu}\partial _{\rho}u_{\sigma},
\end{equation}
here $j_\beta(r_+)$, $j_Q(r_+)$, $j_F(r_+)$ are values of each function at the horizon and same as (\ref{jis}).

For more simplicity, we can obtain that the charged current can be rewritten as an invariant form
\begin{eqnarray}
J^{\mu }_{(1)}=-\kappa  P^{\mu  \nu } \partial _{\nu }\frac{\mu }{T}+\frac{\sigma}{e} u^{\lambda } F ^{\rm ext}{}_{\lambda }{}^{\mu }+\sigma _{B}B^{\mu }+\xi \omega ^{\mu }.
\end{eqnarray}
where
\begin{eqnarray}
\sigma _{B}=\frac{\sqrt{3} \kappa_{cs} Q (3r_{+}^4+2M)}{gMr_{+}^2},~~\xi=\frac{6 \kappa_{cs} Q^2}{M}.
\end{eqnarray}
Note that, the last two terms are related with the anomalous magnetic and vortical effects, and it is also obvious that they are from the Chern-Simons term because the coefficients of the last two terms are proportional to $\kappa_{cs}$.


\section{Conclusion and discussion}
In this paper, we apply the AdS/CFT correspondence to investigate
the property of hydrodynamics of the dual strongly coupling conformal
field by studying the $5$-dimensional solutions of the Maxwell-Gauss-Bonnet
gravity in the bulk. By lifting the parameters of the boosted black brane in the Maxwell-Gauss-Bonnet gravity to functions of boundary coordinates, and then solving for the corresponding correction terms, we finally construct the first order perturbative gravity solution. Base on this perturbative solution, we extract the information of its dual conformal field, such as the stress tensor and conserved current. And we also obtain the shear viscosity $\eta$ , entropy density $s$ , thermal conductivity $\kappa$ and electrical conductivity $\sigma$ of the dual conformal fluid. Through these results, we also find that the Gauss-Bonnet term can affect all these parameters. However, it does not affect the so called Wiedemann-Franz law which relates the thermal with electrical conductivity, while it affects the ratio $\eta/s$ which has been predicted in several previous work. It should be emphasized that usually the generalization of the approach in~\cite{Bhattacharyya:2008jc} to new cases such as including the Maxwell field in the bulk in our case are non-trivial. Because it not only can extract new physics from the bulk such as the conserved current $J^{\mu}$, but also solve
the difficulties in the mathematics which can be seen in the appendix \ref{B} that the coefficients $a_{i}(r)$ couples $j_{i}(r)$ and it is more difficult to solve these coupled coefficients.

Note that, the above parameters $\eta$, $\kappa$, $\sigma$ can also be obtained from the method using the Kubo formula~\cite{Cai:2008ph,Myers:2009ij}. The same results imply that the new approach in~\cite{Bhattacharyya:2008jc} is underlying consistent with the method using the Kubo formula. However, there is a more way to check the consistence for this new approach, because the entropy density can also be obtained from the pressure in (\ref{entr1}) which can be directly compared with (\ref{entr}). Furthermore, as a new approach, it gives a new insight into the AdS/CFT correspondence. It directly show that the stress tensor of the conformal viscous fluid in the filed side is connected with the well known boundary stress tensor (which is used to calculate the conserved quantities such as the energy, momentum) from the counterterm method in the gravity side~\cite{Balasubramanian:1999re,Emparan:1999pm,Mann:1999pc}. Moreover, the explicit expressions of the stress tensor $\tau_{\mu \nu}$ and conserved current $J^{\mu}$ of the dual conformal fluid are naturally deduced in this new approach. Therefore, basing on these explicit expressions, it may obtain some extra terms which are neglected before by using the Kubo formula. For example, if one considers the Chern-Simon term of the Maxwell field in the bulk, the conserved current $J^{\mu}$ following the method in~\cite{Bhattacharyya:2008jc} can contain two additional terms inducing the anomalous magnetic and vortical effects, which has been attracted by many authors~\cite{Erdmenger:2008rm,Banerjee:2008th,Son:2009tf}. And a simple discussion by adding the Chern-Simons term in our case in the Maxwell-Gauss-Bonnet gravity have been given in the above section~V. Therefore, more effects of the Chern-Simon term in the bulk with the modified gravity are interesting to have further study. Moreover, whether there are some new extra terms deduced from the effects of $RF^{2}$ terms of Maxwell filed and more general expressions in~\cite{Myers:2009ij} are also an interesting open question. In addition, one can further use these explicit expressions to model the real hydrodynamical process of QGP through considering its asymptotic late-time behavior~\cite{Bigazzi:2010ku,Kanitscheider:2009as}. Our work showing the effect of Gauss-Bonnet term on the gravity/fluid correspondence may give us some insights to model this real process of QGP.


\section{Acknowledgements}

Y.P Hu thanks Professor Rong-Gen Cai, Dr. Zhang-Yu Nie for useful discussions. This work is
supported partially by grants from NSFC, No. 10975168 and No. 11035008, China.

\appendix

\section{The tensor components of $W_{\mu \nu }$ and $S_{\mu \nu }$}
\label{A} The tensor components of $W_{\mu \nu } = (\text{effect
from correction}) - S_{\mu \nu }$ are
\begin{eqnarray}
W_{vv} &=&\frac{4}{3} \alpha  f(r) f'(r)^2 h'(r) r^5+\frac{4}{3}
\alpha
   f(r)^2 h'(r) f''(r) r^5+\frac{4}{3} \alpha  f(r)^2 f'(r) h''(r)
   r^5+8 \alpha  f(r) h(r) f'(r)^2 r^4 \notag\\
   &&+\frac{52}{3} \alpha  f(r)^2
   f'(r) h'(r) r^4-f(r) f'(r) h'(r) r^4+8 \alpha  f(r)^2 h(r)
   f''(r) r^4-f(r) h(r) f''(r) r^4  \notag\\
   &&+56 \alpha  f(r)^2 h(r) f'(r)
   r^3-7 f(r) h(r) f'(r) r^3+8 \alpha  f(r)^3 h'(r) r^3-2 f(r)^2
   h'(r) r^3+32 \alpha  f(r)^3 h(r) r^2 \notag\\
   &&-8 f(r)^2 h(r) r^2+2 \alpha
    k(r) f'(r)^2+4 \alpha  f(r) f'(r) k'(r)+4 \alpha  f(r) k(r)
   f''(r)-\frac{1}{2} k(r) f''(r) \notag\\
   &&+2 \alpha  f(r)^2
   k''(r)-\frac{1}{2} f(r) k''(r)+\frac{4 Q f(r)
   a_{v}'(r)}{\sqrt{3} \text{g} r}+\frac{12 \alpha  f(r)
   k(r) f'(r)}{r}-\frac{7 k(r) f'(r)}{2 r} \notag\\
   &&-\frac{2 \alpha  f(r)^2
   k'(r)}{r}+\frac{f(r) k'(r)}{2 r}+\frac{8 \alpha  f(r)^2
   k(r)}{r^2}-\frac{4 f(r) k(r)}{r^2}+\frac{4 k(r)}{r^2}+\frac{8
   Q^2 f(r) h(r)}{r^4}\notag\\
   &&+\frac{4 Q^2 k(r)}{r^8}-S_{vv}, \notag\\
W_{vr} &=&-\frac{4}{3} \alpha  f'(r)^2 h'(r) r^3-\frac{4}{3} \alpha
f(r)
   h'(r) f''(r) r^3-\frac{4}{3} \alpha  f(r) f'(r) h''(r) r^3-6
   \alpha  h(r) f'(r)^2 r^2  \notag\\
   &&-\frac{52}{3} \alpha  f(r) f'(r) h'(r)
   r^2+f'(r) h'(r) r^2-6 \alpha  f(r) h(r) f''(r) r^2+\frac{1}{2}
   h(r) f''(r) r^2 \notag\\
   &&-42 \alpha  f(r) h(r) f'(r) r+\frac{7}{2} h(r)
   f'(r) r-8 \alpha  f(r)^2 h'(r) r+2 f(r) h'(r) r-24 \alpha
   f(r)^2 h(r)\notag\\
   &&+4 f(r) h(r)+4 h(r)-\frac{4 \alpha  f'(r)
   k'(r)}{r^2}-\frac{2 \alpha  k(r) f''(r)}{r^2}-\frac{2 \alpha
   f(r) k''(r)}{r^2}+\frac{k''(r)}{2 r^2}-\frac{4 Q
   a_{v}'(r)}{\sqrt{3} \text{g} r^3}\notag\\
   &&+\frac{2 \alpha  k(r)
   f'(r)}{r^3}+\frac{2 \alpha  f(r) k'(r)}{r^3}-\frac{k'(r)}{2
   r^3}-\frac{4 Q^2 h(r)}{r^6}-S_{vr},\notag\\
W_{vx}&=&-\frac{2 j_{x}(r) Q^2}{r^8}+\frac{\sqrt{3} f(r)a_{x}'(r)Q}{\text{g} r}
   +\frac{8 \alpha  f(r)^2
   j_{x}(r)}{r^2}-\frac{4 f(r) j_{x}(r)}{r^2}+\frac{4
   j_{x}(r)}{r^2}-\frac{4 \alpha  f(r) j_{x}(r)
   f'(r)}{r}\notag\\
   &&-\frac{j_{x}(r) f'(r)}{r}-\frac{6 \alpha  f(r)^2
   j_{x}'(r)}{r}+\frac{3 f(r) j_{x}'(r)}{2 r}+2 \alpha
   f(r) f'(r) j_{x}'(r)+2 \alpha  f(r)^2
   j_{x}''(r)\notag\\
   &&-\frac{1}{2} f(r) j_{x}''(r)-S_{vx},\notag\\
W_{rr}&=&-\frac{20 \alpha  f(r) h'(r)}{r}+\frac{5 h'(r)}{r}-4 \alpha
 f(r)h''(r)+h''(r)-S_{rr}, \notag\\
W_{rx}&=&-\frac{\sqrt{3} Q a_{x}'(r)}{\text{g} r^3}+\frac{8
\alpha
   j_{x}(r) f'(r)}{r^3}+\frac{6 \alpha  f(r)
   j_{x}'(r)}{r^3}-\frac{2 \alpha  f'(r)
   j_{x}'(r)}{r^2}-\frac{3 j_{x}'(r)}{2 r^3}-\frac{2
   \alpha  f(r) j_{x}''(r)}{r^2}\notag\\
   &&+\frac{j_{x}''(r)}{2 r^2}-S_{rx},\notag\\
W_{ij}\delta ^{ij} &=&-\frac{12 \alpha  f(r) f'(r) h'(r)
r^4}{\ell_{c}^2}+\frac{f'(r) h'(r)
   r^4}{\ell_{c}^2}-\frac{4 \alpha  f(r)^2 h''(r) r^4}{\ell_{c}^2}+\frac{f(r)
   h''(r) r^4}{\ell_{c}^2} \notag\\
   &&-\frac{56 \alpha  f(r) h(r) f'(r)
   r^3}{\ell_{c}^2}+\frac{8 h(r) f'(r) r^3}{\ell_{c}^2}-\frac{44 \alpha  f(r)^2
   h'(r) r^3}{\ell_{c}^2}+\frac{11 f(r) h'(r) r^3}{\ell_{c}^2}\notag\\
   &&-\frac{112 \alpha
   f(r)^2 h(r) r^2}{\ell_{c}^2}+\frac{32 f(r) h(r) r^2}{\ell_{c}^2}-\frac{8 h(r)
   r^2}{\ell_{c}^2}+\frac{2 \sqrt{3} Q a_{v}'(r)}{\text{g} {\ell_{c}^2}
   r}-\frac{12 \alpha  k(r) f'(r)}{{\ell_{c}^2} r}\notag\\
   &&-\frac{12 \alpha  f(r)
   k'(r)}{{\ell_{c}^2} r}+\frac{3 k'(r)}{{\ell_{c}^2} r}+\frac{16 Q^2 h(r)}{{\ell_{c}^2} r^4}-S_{ij}\delta ^{ij}, \notag\\
W_{ij} &=&\frac{1}{3}\delta _{ij}(\delta ^{kl}W_{kl})-\frac{1}{2r \ell_{c}^2}
(r^{5}f(1-4\alpha f-2\alpha rf^{\prime })\alpha _{ij}^{\prime
})^{\prime} -S_{ij}+\frac{1}{3}\delta _{ij}(\delta ^{kl}S_{kl}),\notag
\end{eqnarray}
where the first order source terms are
\begin{eqnarray}
S_{vv} &=&\frac{4}{3} \alpha \partial_{i} \beta_{i} f(r) f''(r)
r^3-\frac{1}{2}  \partial_{i} \beta_{i}f'(r) r^2
   +\frac{26}{3}  \alpha \partial_{i} \beta_{i} f(r) f'(r) r^2+4 \alpha \partial_{i} \beta_{i} f(r)^2 r+\frac{3}{2}
   \partial_{v}{M} C_{1}(r) r \notag\\
   &&+\frac{3}{2} \partial_{v}{Q} C_{2}(r)
   r-\partial_{i} \beta_{i} f(r) r -6 \partial_{v}{M} \alpha  C_{1}(r)
   f(r) r-6 \partial_{v}{Q} \alpha C_{2}(r) f(r) r, ~\notag\\
S_{vr} &=&-\frac{4 \alpha  f(r) \partial_{i} \beta_{i}}{r}-\frac{20}{3} \alpha f'(r) \partial_{i} \beta_{i}-\frac{4}{3} r \alpha  f''(r) \partial_{i} \beta_{i}+\frac{\partial_{i} \beta_{i}}{r}, \notag\\
S_{vi} &=&\frac{3 \partial_{v} \beta_{i} Q^2}{ r^5}-\frac{\sqrt{3}
F_{vi}^{ext} e
   Q}{\text{g} r^3}+\frac{3 \partial_{i}Q Q}{ r^5}-6 r \alpha
   \partial_{v} \beta_{i} f(r)^2+\frac{3}{2} \partial_{i}M r
   C_{1}(r)\notag\\
   &&+\frac{3}{2} \partial_{i}Q r C_{2}(r)+\frac{3}{2} r
   \partial_{v} \beta_{i} f(r)-6 \partial_{i}M r \alpha  C_{1}(r)
   f(r)-6 \partial_{i}Q r \alpha  C_{2}(r) f(r)+\frac{1}{2}
   \partial_{i}M r^2 C_{1}'(r)\notag\\
   &&-2 \partial_{i}M r^2 \alpha  f(r)
   C_{1}'(r)+\frac{1}{2} \partial_{i}Q r^2 C_{2}'(r)-2
   \partial_{i}Q r^2 \alpha  f(r) C_{2}'(r)+\frac{1}{2} r^2
   \partial_{v} \beta_{i} f'(r)\notag\\
   &&-2 \partial_{i}M r^2 \alpha  C_{1}(r)
   f'(r)-2 \partial_{i}Q r^2 \alpha  C_{2}(r) f'(r)-4 r^2 \alpha
   \partial_{v} \beta_{i} f(r) f'(r),\notag\\
S_{rr} &=&0, \notag\\
S_{ri} &=&\frac{6 \alpha  f(r) \partial_{v} \beta_{i}}{r}+2 \alpha f'(r) \partial_{v} \beta_{i}-\frac{3 \partial_{v}\beta_{i}}{2 r},\notag\\
S_{xy} &=&(-\frac{\alpha  f''(r) r^3}{\ell_{c}^2}-\frac{6 \alpha  f'(r) r^2}{\ell_{c}^2}+\frac{3 r}{2 \ell_{c}^2}-\frac{6 \alpha
   f(r) r}{\ell_{c}^2})(\partial_{x}\beta_{y}+\partial_{y}\beta_{x}),\notag\\
S_{xx} &=&-\frac{4 \alpha  \partial_{x} \beta_{x} f''(r) r^3}{3\ell_{c}^2}+\frac{2
   \alpha  \partial_{y} \beta_{y} f''(r) r^3}{3 \ell_{c}^2}+\frac{2 \alpha
   \partial_{z} \beta_{z} f''(r) r^3}{3 \ell_{c}^2}-\frac{32 \alpha
   \partial_{x} \beta_{x} f'(r) r^2}{3 \ell_{c}^2}+\frac{4 \alpha
   \partial_{y} \beta_{y} f'(r) r^2}{3 \ell_{c}^2}\notag\\
   &&+\frac{4 \alpha
   \partial_{z} \beta_{z} f'(r) r^2}{3 \ell_{c}^2}+\frac{4 \partial_{x} \beta_{x}
   r}{\ell_{c}^2}+\frac{\partial_{y} \beta_{y} r}{\ell_{c}^2}+\frac{\partial_{z} \beta_{z}
   r}{\ell_{c}^2}-\frac{16 \alpha \partial_{x} \beta_{x} f(r) r}{\ell_{c}^2}-\frac{4
   \alpha  \partial_{y} \beta_{y} f(r) r}{\ell_{c}^2}-\frac{4 \alpha \partial_{z} \beta_{z} f(r) r}{\ell_{c}^2},\notag
\end{eqnarray}
and $S_{yy}$ (or $S_{zz}$) can be just replaced the index $x$ in $S_{xx}$ into $y$ (or $z$), and
\begin{equation}
C_{1}(r)=\frac{\partial f(r)}{\partial M},~~C_{2}(r)=\frac{\partial f(r)}{\partial Q}.
\end{equation}

\section{The exact form of $j_i(r)$ and $a_i(r)$} \label{B}
In this appendix we will deduce the $j_i(r)$ and $a_i(r)$ in detail
from coupled differential equations (\ref{appeq1}). Like \cite{Hur:2008tq}, a second order differential
equation of $j_i(r)$ can be obtained
\begin{eqnarray}
j_i{}^{\prime\prime }(r)-(\frac{3}{r}-\frac{4\alpha
f'(r)}{-1+4\alpha f(r)}) j_i'(r)-(\frac{-12 Q^2+16r^7 \alpha
f(r)f'(r)} {r^8 f(r)(-1+4\alpha f(r))} )j_i(r)  = \zeta_i(r),
\label{appeqj}
\end{eqnarray}
where
\begin{equation}
\zeta_i(r)\equiv\big(-\frac{12 Q^2}{r^4 f(r)} \frac{j_i\left(r_+\right)}{r_+^4}+2 r^2 S_{r i}(r)+\frac{2 \sqrt{3} Q}{g r^4 f(r)} \int _{r_+}^rdx x S_i(x))/(1-4\alpha f(r)).
\end{equation}

And then we can write out a particular solution to (\ref{appeqj})
\begin{eqnarray}
j_P(r)=b_1(r) j_{H_1}(r)+b_2(r) j_{H_2}(r),
\end{eqnarray}
where
\begin{eqnarray}
b_1(r)&=&-\int _r^{\infty }dx \frac{j_{H_2}(x) \zeta _i(x)(1-4\alpha f(x))}{x^3}, \\
b_2(r)&=&r^3 (2-\frac{1}{\ell_{c}^2})\partial _v\beta _i+\int
_r^{\infty }dx \left(\frac{j_{H_1}(x) \zeta _i(x)(1-4\alpha
f(x))}{x^3}+3 x^2 (2-\frac{1}{\ell_{c}^2})\partial _v\beta_i\right).
\end{eqnarray}
and
\begin{eqnarray}
j_{H_1}(r)&=&r^4 f(r), \\
j_{H_2}(r)&=&j_{H_1}(r) \int _r^{\infty }\frac{x^3
dx}{j_{H_1}(x){}^2}=r^4 f(r) \int _r^{\infty }\frac{dx}{x^5 f(x)^2
(1-4\alpha f(x))}.
\end{eqnarray}
are two linearly independent homogeneous solutions of (\ref{appeqj}). Here, the $3 x^2 \partial _v\beta _i$ term is added to cancel the divergence of the integral.
With the above formulas, the general solution to (\ref{appeqj}) can be represent as
\begin{eqnarray}
j_i(r)=j_P(r)+C_3 j_{H_1}(r)+C_4 j_{H_2}(r).
\end{eqnarray}
Where the integration constant $C_3$ and $C_4$ can be determined by the asymptotic behavior of $j_i(r)$
\begin{eqnarray}
j_i(r)=\frac{C_3 r^4}{\ell_{c}^2}+r^3 \partial _v\beta _i+\frac{2C_3M(-1+\sqrt{1-8 \alpha})+ C_4 \alpha}
{-1+8 \alpha+\sqrt{1-8 \alpha}}+O\left(\frac{1}{r}\right).\label{appjsol1}
\end{eqnarray}
Since the first term in (\ref{appjsol1}) is non-normalizable, $C_3$ is forced to zero. The other integration constant $C_4$ can be set to zero by demanding $u^\mu T^{(n)}_{\mu\nu}=0$. In summary,
\begin{eqnarray}
j_i(r)&=&-r^4 f(r)\int _r^{\infty }dx x f(x)(1-4 \alpha f(x)) \zeta _i(x) \int _x^{\infty }\frac{dy}{y^5 f(y)^2 (1-4 \alpha f(y))} \nonumber\\
&& +r^4 f(r) \left(\int _r^{\infty }\frac{dx}{x^5 f(x)^2 (1-4 \alpha f(x))}\right) \Big(r^3(2-\frac{1}{\ell_{c}^2}) \partial _v\beta _i \nonumber\\
&&+\int _r^{\infty }dx \left[x f(x)(1-4 \alpha f(x)) \zeta _i(x)+3 x^2 (2-\frac{1}{\ell_{c}^2})\partial _v\beta _i\right]\Big). \label{appjsol}
\end{eqnarray}

And asymptotic behavior of $j_i(r)$ can be present as:
\begin{eqnarray}
j_i(r) &\approx& r^3 \partial _v\beta _i-\frac{8 \sqrt{3} Q \alpha}{5 r (-1+8 \alpha +\sqrt{1-8 \alpha})} \frac{e}{g}F_{v i}^{\text{\rm ext}}+\frac{4 \alpha}{r^2(-1+8 \alpha +\sqrt{1-8 \alpha})} \nonumber\\
&&\left(-Q^2 \frac{j_i\left(r_+\right)}{r_+^4}-\frac{Q}{2 r_+} \left(\partial _iQ+Q \partial _v\beta _i\right)\right. \nonumber\\
&& \left.+\frac{r_+ Q}{2 \sqrt{3}} \frac{e}{g}F_{v i}^{\text{\rm
ext}}\right).
\end{eqnarray}
To obtain $j_i(r_+)$, we take $r\to r_+$ limit to (\ref{appjsol})
and get
\begin{eqnarray}\label{j at rp}
\frac{j_i\left(r_+\right)}{r_+^4}&=&\frac{2 r_+^2 \left(2 r_+^4+r_+^2 r_-^2+r_-^4\right) \partial _v\beta _i-Q \left(\partial _iQ+Q \partial _v\beta _i\right)-\sqrt{3} r_+^2 Q \frac{e}{g}F_{v i}^{\text{\rm ext}}}{6 r_+ Q^2-2 r_+^3 \left(-2 r_+^4+r_+^2 r_-^2+r_-^4\right)} \nonumber\\
&=&\frac{2 \left(2 r_+^6+Q^2\right) \partial _v\beta _i-Q \left(\partial _iQ+Q \partial _v\beta _i\right)-\sqrt{3} r_+^2 Q \frac{e}{g}F_{v i}^{\text{\rm ext}}}{8 M r_+^3}.
\end{eqnarray}
From the definition
\begin{eqnarray}
j_i(r) &=& j_\beta(r) \partial_v \beta_i + j_Q(r) (\partial_i Q+Q\partial_v \beta_i) +j_F(r) F^{\rm ext}_{vi},
\end{eqnarray}
we can read off the result (\ref{jis}).

In addition, integrating the second equation in (\ref{appeq1}) from $r=r_+$ to $r=\infty$, we get
\begin{eqnarray}
r^3 f(r) a_i'(r)-2 \sqrt{3} g Q
\left(\frac{j_i(r)}{r^4}-\frac{j_i\left(r_+\right)}{r_+^4}\right)=\int
_{r_+}^rdx x S_i(x), \label{appeq2i}
\end{eqnarray}
Having the expression of $j_i(r)$ and $j_i(r_{+})$, $a_i(r)$ is obtained by
integrating (\ref{appeq2i})
\begin{eqnarray}
a_i(r) &=& \int_\infty^r dx
\frac{1}{x^3f(x)}\left(2\sqrt{3}g Q(\frac{j_{i}(x)}{x^4}-\frac{j_i(r_+)}{r_{+}^4})+
\int_{r_+}^x dy y S_{i}(y)\right),
\end{eqnarray}
where the gauge that make $a_i(r)$ vanishes at infinity is applied.
At last asymptotic behavior of $a_i(r)$ can be present as:
\begin{eqnarray}
a_i(r) &\approx& \frac{\ell_{c}^2}{r} eF_{v i}^{\text{\rm
ext}}+\frac{\ell_{c}^2}{r^2} \left(\sqrt{3} gQ
\frac{j_i\left(r_+\right)}{r_+^4}+\frac{\sqrt{3}g}{2 r_+}
\left(\partial _iQ+Q \partial _v\beta _i\right)-\frac{r_+}{2} eF_{v
i}^{\text{\rm ext}}\right).
\end{eqnarray}


\end{document}